\documentclass[10pt]{article}
\usepackage{stmaryrd}
\usepackage{amsfonts}
\usepackage{amsmath}
\usepackage{latexsym,amssymb,amsfonts,amsbsy, amsthm}
\usepackage[usenames]{color}
\usepackage{xspace,colortbl}
\newcommand{\md}{\mathrm{d}}

\newtheorem{thm}{Theorem}[section]

\numberwithin{equation}{section} 

\begin{document}

\title{\textbf{Solutions of Jimbo-Miwa Equation and Konopelchenko-Dubrovsky Equations}}

\author{Bintao Cao\footnote{Institute of Mathematics, Academy of Mathematics and Systems Science,
Chinese Academy of Sciences, Beijing 100190, P. R. China,
  caobintao@amss.ac.cn} \ }
\date{}
\maketitle

\begin{abstract}
\noindent The Jimbo-Miwa equation is the second equation in the well
known KP hierarchy of integrable systems, which is used to describe
certain interesting (3+1)-dimensional waves in physics but  not pass
any of the conventional integrability tests. The
Konopelchenko-Dubrovsky equations arose in physics in connection
with the nonlinear weaves with a weak dispersion.  In this paper, we
obtain two families of explicit exact solutions with multiple
parameter functions for these equations by using Xu's stable-range
method and our  logarithmic generalization of the stable-range
method. These parameter functions make our solutions more applicable
to related practical models and boundary value problems.
\end{abstract}

\noindent {\sl Keywords:} Jimbo-Miwa; Konopelchenko-Dubrovsky;
Stable-range; Logarithmic stable-range.\

\vskip 0.2cm

\noindent {\sl AMS Subject Classification (2000):} 35Q51, 35C10,
35C15. \

\renewcommand{\theequation}{\thesection.\arabic{equation}}
\setcounter{equation}{0}
\section{Introduction}
 Jimbo and Miwa  \cite{J-M} (1983) first studied the following
 nonlinear partial differential equation:
\begin{equation}\label{j-m}
W_{xxxy}+3W_{xy}W_x+3W_yW_{xx}+2W_{yt}-3W_{xz}=0,
\end{equation}
as the second equation  in the well known KP hierarchy of integrable
systems. The equation is used to describe certain interesting
(3+1)-dimensional waves in physics but not pass any of the
conventional integrability tests \cite{D-G-R-W}. One of the
important features is that the equation has soliton solutions. The
space of $\tau$- functions for this hierarchy, given by Jimbo and
Miwa \cite{J-M}(1983), is the orbit of the vacuum vector for the
Fock representation of the Lie algebra $gl(\infty)$. Dorizzi,
Grammaticos, Ramani and  Winternitz \cite{D-G-R-W} (1986) calculated
 Lie symmetries of \eqref{j-m} in terms of Lie algebra. They showed that
the algebra is infinite dimensional, but does not have the
Kac-Moody-Virasoro algebra structure. Rubin and Winternitz
\cite{R-W} (1990) found that the joint symmetry algebra of the
system of the first two equations in KP hierarchy have a
Kac-Moody-Virasoro algebra structure. The generalized $W_{\infty}$
symmetry algebra of these two equations were found by Lou and Weng
\cite{L-W} (1995). Hong and Oh \cite{H-O} (2000) got a class of
solitary wave solutions of \eqref{j-m} by generalizing the tanh
method. Fan \cite{F} (2003) obtained a line solitary wave solution,
a Jocobi doubly periodic solution and a Weierstrass periodic
solution by using modified tanh method.  Abdou \cite{A} (2008) found
some generalized solitary  wave solutions and periodic solutions by
the exp-function method.

The  equations
\begin{subequations}
\begin{align}
&u_t-u_{xxx}-6buu_x+{3\over 2}a^2u^2u_x-3v_y+3au_xv=0\label{k-da}\\
&u_y=v_x\label{k-db}
\end{align}
\end{subequations}
 were introduced by
Konopelchenko and  Dubrovsky \cite{K-D} (1984) in connection with
the nonlinear weaves with a weak dispersion, where $a$ and $b$ are
real constants. These equations can be represented as the
commutativity $[L,T]=0$ of certain differential operators $L$ and
$T$ \cite{K-D}. The system is the two dimensional generalization of
the well-known Gardner equation, KP equation (the first equation of
the KP hierarchy) and the modified KP equation.  Maccari \cite{M}
(1999) derived an integrable Davey-Stewartson-type equation from
\eqref{k-da} and \eqref{k-db}. H. Zhi \cite{Zhi} (2008) found the
symmetry group of this system. To solve the Konopelchenko-Dubrovsky
equations, various methods have been proposed, such as the standard
truncated Painlev\'{e} analysis \cite{L-L-W}, the tanh method and
its generalizations \cite{X-L-Z}\cite{Zha2}\cite{Zhi}, the
generalized F-expansion method \cite{W-Z}\cite{Z-X}\cite{Zha1}, the
extended Riccati equation rational expansion method \cite{S-Z1},
exp-function method \cite{A}, the tanh-sech method, the cosh-sinh
method, the exponential functions method \cite{W} and the homotopy
perturbation method \cite{S-Z2}.

Most of the above existing exact explicit solutions of the
Jimbo-Miwa equation and the Konopelchenko-Dubrovsky equations are
traveling-wave-type solutions and their slightly generalizations.
These solutions do not fully reflect the features of these nonlinear
partial differential equations. It is desirable to find new exact
explicit solutions that capture more features of these equations.

Using certain finite-dimensional stable range of the nonlinear term,
Xu \cite{X1} found a family of exact solutions with seven parameter
functions for the equation of nonstationary transonic gas flows,
which blow up on a moving line. Moreover, he \cite{X2} solved the
short wave equation and the Khokhlov-Zabolotskaya equations by the
same method and obtained certain interesting singular and smooth
explicit exact solutions with multiple parameter functions.

In this paper, we find two families of explicit exact solutions with
multiple parameter functions for the Jimbo-Miwa equation and the
Konopelchenko-Dubrovsky equations by using Xu's stable-range method
and our logarithmic generalization of the stable-range method,
motivated from the standard truncated Painlev\'{e} analysis, as in
\cite{L-L-W}. The first family of solutions are polynomial in the
variable $x$ or $y$. The second family solutions are not polynomial
in any variable. They are logarithms of the functions that are
polynomial either in $x$ or in $y$. Our solutions in general are not
traveling-wave-type solutions. Their multiple-parameter-function
feature makes them more applicable to related practical models and
boundary value problems.

In Section 2, we find exact solutions of the  the Jimbo-Miwa
equation. The Konopelchenko-Dubrovsky equations are solved in
Section 3.

\section{Solutions of the Jimo-Miwa Equation}
\subsection{Stable-Range Approach}
We assume that
\begin{equation}\label{2.1}
W=\sum_{m=0}^nA_m(y,z,t)x^m
\end{equation} is a solution of the Jimbo-Miwa equation \eqref{j-m}.
First we consider the case $n\leq2$, i.e.
\begin{equation}\label{2.2}
W=A_2x^2+A_1x+A_0.
\end{equation} Note that
\begin{equation}\label{2.3}
W_x=2A_2x+A_1,\ \ W_{xx}=2A_2,\ \ W_{xxx}=0,
\end{equation}
\begin{equation}\label{2.4}
W_y=A_{2y}x^2+A_{1y}x+A_{0y},\ \ W_{xy}=2A_{2y}x+A_{1y},
\end{equation}
and
\begin{equation}\label{2.5}
W_{xz}=2A_{2z}x+A_{1z},\ \ W_{yt}=A_{2yt}x^2+A_{1yt}x+A_{0yt}.
\end{equation}

Substituting \eqref{2.2}-\eqref{2.5} into \eqref{j-m}, we get
\begin{eqnarray}\label{2.6}
&&3(2A_{2y}x+A_{1y})(2A_{2}x+A_1)+3(A_{2y}x^2+A_{1y}x+A_{0y})\cdot2A_2\nonumber\\
&+&2(A_{2yt}x^2+A_{1yt}x+A_{0yt})-3(2A_{2z}x+A_{1z})=0.
\end{eqnarray}
Thus
\begin{subequations}
\begin{align}
9A_2A_{2y}+A_{2yt}=0,\label{2.7a}\\
3A_1A_{2y}+6A_{1y}A_2+A_{1yt}-3A_{2z}=0,\label{2.7b}\\
3A_1A_{1y}+6A_2A_{0y}+2A_{0yt}-3A_{1z}=0.\label{2.7c}
\end{align}
\end{subequations}

Observe that
\begin{equation}\label{2.8}
A_2=\alpha_t(t,z)
\end{equation}
and
\begin{equation}\label{2.9}
A_2=({9\over2}t+\beta(y,z))^{-1}
\end{equation}
are solutions of \eqref{2.7a}, where $\alpha$ and $\beta$ are
arbitrary differential functions. Throughout this paper, the
indefinite integration means an antiderivative of the integrand with
respect to the integral variable.
 Substituting \eqref{2.8} into
\eqref{2.7b}, we get
\begin{equation}\label{2.10}
6\alpha_t A_{1y}+A_{1yt}-3\alpha_{tz}=0.
\end{equation} It implies
\begin{eqnarray}\label{2.11}
A_1=e^{-6\alpha}\left(\gamma(y,z) +3y\int\alpha_{tz} e^{6\alpha}\md
t\right) +\rho(z,t),
\end{eqnarray}
where $\gamma(y,z)$ and $\rho(z,t)$ are arbitrary functions.
Similarly, we get
\begin{eqnarray}\label{2.12}
A_0=e^{-3\alpha}\left(\eta(y,z)+{3\over2}\iint(A_{1z}-A_1A_{1y})e^{3\alpha}\md
t\md y\right) +\zeta(z,t),
\end{eqnarray}
where $\eta(y,z)$ and $\zeta(z,t)$ are arbitrary functions.

\begin{thm} For arbitrary functions $\alpha(t,z)$, $\eta(y,z)$, $\gamma(y,z)$, $\rho(z,t)$ and
$\zeta(z,t)$, we have the  solution
\begin{eqnarray}\label{p1}
W&=&\alpha_t(t,z)x^2+A_1x+e^{-3\alpha}(\eta(y,z)\nonumber\\
&&+{3\over2}\int\int(A_{1z}-A_1A_{1y})e^{3\alpha }\md t\md
y)+\zeta(z,t)
\end{eqnarray} of the Jimbo-Miwa equation \eqref{j-m}, where $A_1$ is given in \eqref{2.11}.
\end{thm}

Next we deal with  $A_2=({9\over2}t+\beta(y,z))^{-1}$. Assume
\begin{equation}\label{2.13}
A_1=\sum_{n\in \mathbb{Z}}B_n(y,z)({9\over2}t+\beta(y,z))^n.
\end{equation}
Then
\begin{eqnarray}\label{2.14}
A_{1y}=\sum_{n\in
\mathbb{Z}}(B_{ny}+(n+1)B_{n+1}\beta_y)({9\over2}t+\beta(y,z))^n
\end{eqnarray}
and
\begin{eqnarray}\label{2.15}
A_{1yt}=\sum_{n\in
\mathbb{Z}}{9\over2}(n+1)(B_{(n+1)y}+(n+2)B_{n+2}\beta_y)({9\over2}t+\beta(y,z))^n.
\end{eqnarray}
Thus by \eqref{2.7b}, we have
\begin{eqnarray}\label{2.16}
&&\sum_{n\in \mathbb{Z}}({{(n+3)(3n+4)}\over
2}B_{n+2}\beta_y+{{3n+7}\over2}B_{(n+1)y})({9\over2}t+\beta(y,z))^n\nonumber\\
&=&\frac{-\beta_z}{({9\over2}t+\beta(y,z))^2}.
\end{eqnarray}
Hence
\begin{subequations}
\begin{align}
-B_0\beta_y+{1\over2}B_{(-1)y}&=-\beta_z, & n&=-2\label{2.17a}\\
{{(n+3)(3n+4)}\over 2}B_{n+2}\beta_y+{{3n+7}\over2}B_{(n+1)y}&=0. &
n&\neq-2\label{2.17b}
\end{align}
\end{subequations}
Let $n=-3$ in \eqref{2.17b}. We have  $B_{(-2)y}=0$. Then
\begin{equation}\label{2.18}
B_{-2}=\gamma_{-2}(z),
\end{equation}
where $\gamma_{-2}$ is an arbitrary function. Thus, by \eqref{2.18},
\begin{eqnarray}\label{2.19}
B_{-l-2}=\sum_{m=0}^l\frac{3l+5}{3m+5}{l\choose
m}\gamma_{-2-m}(z)\beta^{l-m},
\end{eqnarray}
where $\gamma_{-2-m}(z)$ are arbitrary functions.

If $\beta_y=0$, then
\begin{equation}\label{2.20}
B_{l-1}=\gamma_{l-1}(z),\ \ \mbox{for}\ l\geq1,
\end{equation} and
\begin{equation}\label{2.21}
B_{-1}=-2\beta_zy+\gamma_{-1}(z).
\end{equation} It is not interesting.
Thus, we assume that $\beta_y\neq0$. Let $B_n=\gamma_n(z)$ for $n\in
\mathbb{N}$. Then $B_{(n+1)}=0$, and we have that
\begin{eqnarray}\label{2.22}
B_m=\sum_{r=0}^{n-m}(-1)^{n-m+r}\frac{3m+1}{3(n-r)+1}{n+1-r \choose
m+1}\beta^{n-m-r}\gamma_{n-r}(z)
\end{eqnarray}
for $0\leq m\leq n.$
\begin{eqnarray}\label{2.23}
B_{-1}&=&2\int
B_0\beta_y\md y-2\int\beta_z\md y+\gamma_{-1}(z)\nonumber\\
&=&2\sum_{r=0}^n(-1)^{n+r}\frac{1}{3(n-r)+1}\gamma_{n-r}(z)\beta^{n-r+1}-2\int\beta_z\md y+\gamma_{-1}(z).\nonumber\\
\end{eqnarray}
Hence, we get that
\begin{eqnarray}\label{2.24}
A_0&=&\int({9\over 2}t+\beta)^{-{2\over3}}\eta(y,z)\md
y\nonumber\\&&+\int({9\over 2}t+\beta)^{-{2\over3}}(\int({9\over
2}t+\beta)^{2\over3}(A_{2z}-A_1A_1y)\md t)\md y +\zeta(z,t).\nonumber\\
\end{eqnarray}

\begin{thm}
For arbitrary functions $\beta(y,z)$, $\gamma_s(z)$, $\eta(y,z)$ and
$\zeta(z,t)$, the function
\begin{eqnarray}\label{p3}
W&=&({9\over2}t+\beta(y,z))^{-1}x^2+A_1x+\int({9\over
2}t+\beta)^{-{2\over3}}\eta(y,z)\md y\nonumber\\
&&+\int({9\over 2}t+\beta)^{-{2\over3}}(\int({9\over
2}t+\beta)^{2\over3}(-\beta_z({9\over 2}t+\beta)^{-2}-A_1A_{1y})\md
t)\md y\nonumber\\&& +\zeta(z,t)
\end{eqnarray} is a solution of the Jimbo-Miwa equation \eqref{j-m}, where
\begin{eqnarray}\label{p4}
A_1&=&(\sum_{r=0}^n(-1)^{n+r}{2\gamma_{n-r}(z)\over
3(n-r)+1}\beta^{n-r+1}-2\int \beta_z\md y
+\gamma_{-1}(z))({9\over 2}t+\beta)^{-1}\nonumber\\
&&+\sum_{m=0}^n\sum_{r=0}^{n-m}(-1)^{n-m+r}{3m+1\over
3(n-r)+1}{n+1-r\choose m+1}\nonumber\\
&&\times\beta^{n-m-r}\gamma_{n-r}(z)({9\over 2}t+\beta)^m.\nonumber\\
\end{eqnarray}
\end{thm}

 Now we consider the case $n\geq3$ in \eqref{2.1}. In this case,
\begin{equation}\label{2.25}
A_m=A_m(z,t)\ \ \ \ \ \ \ \mbox{for}\ \ \ m=2,\ldots,n.
\end{equation}
We have
\begin{eqnarray}\label{2.26}
&&3\sum_{m=0}^{n-1}(m+1)^2A_{1y}A_{m+1}x^m+3\sum_{m=0}^{n-2}(m+2)(m+1)A_{0y}A_{m+2}x^m\nonumber\\
&-&3\sum_{m=0}^{n-1}(m+1)A_{(m+1)z}x^m+2A_{1yt}x+2A_{0yt}=0,
\end{eqnarray}
by \eqref{j-m}. Note that
\begin{equation}\label{2.27}
A_{1y}=\frac{A_{nz}}{nA_n},
\end{equation}
which implies
\begin{equation}\label{2.28}
A_1=\frac{A_{nz}}{nA_n}y+\eta(z,t),
\end{equation}
and
\begin{subequations}
\begin{align}
A_{(m+1)z}&=(m+1)A_{1y}A_{m+1}+(m+2)A_{0y}A_{m+2},\ \ \mbox{for}\ m=2,\ldots,n,\label{2.29a}\\
A_{2z}&=2A_{1y}A_2+3A_{0y}A_3+{1\over3}A_{1yt},\label{2.29b}\\
A_{1z}&=A_{1y}A_1+2A_{0y}A_2+{2\over3}A_{0yt},\label{2.29c}
\end{align}
\end{subequations} where $A_{n+r}=0$ for $r>0$.
Then
\begin{equation}\label{2.30}
(\frac{A_{nz}}{nA_n})^2-(\frac{A_{nz}}{nA_n})_z=0.
\end{equation}
So, we get that
\begin{equation}\label{2.31}
A_n=\gamma_n(t)(-z+g(t))^{-n},
\end{equation}
where $\gamma_n(t)$ and $g(t)$ are arbitrary functions. By
induction, we obtain that
\begin{eqnarray}\label{2.32}
A_{n-m}={\prod_{s=0}^{m-1}(n-s)\over
(-z+g)^{n-m}}\sum_{s=0}^m\gamma_{n-s}(t){(\int\frac{\varphi}{-z+g}\md
z)^{m-s}\over (m-s)!}
\end{eqnarray} for $m=0,\ldots,n-3$,
where $\varphi=A_{0y}$, and $\gamma_{n-s}(t)$ are arbitrary
functions. By \eqref{2.29b},
\begin{eqnarray}\label{2.33}
A_2&=&(\prod_{s=0}^{n-3}(n-s))(-z+g)^{-2}\sum_{s=0}^{n-2}\gamma_{n-s}{(\int\frac{\varphi}{-z+g}\md
z)^{n-2-s}\over (n-2-s)!}\nonumber\\
&&-\frac{g_tz}{3}(-z+g)^{-2}
\end{eqnarray}
Moreover, by \eqref{2.29c}, we have
\begin{equation}\label{2.34}
\eta={1\over -z+g}(2\int(\varphi A_2+{\varphi_t\over3})(-z+g)\md
z+h(t)),
\end{equation}
where $h(t)$ ia an arbitrary function. Then by \eqref{2.28}, we can
get the explicit form of $A_1$. Moreover, $A_{0y}=\varphi$.
Integrate the function $\varphi(z,t)$, we obtain that
\begin{equation}\label{2.34.1}
A_0=y\varphi(z,t)+f(z,t),
\end{equation} where $f$ is an arbitrary function.
\begin{thm} Let The $n>2$ be an integer, and let $g(t)$,
$\gamma_s(t)$, $\varphi(z,t)$, $f(z,t)$ and $h(t)$ be arbitrary
functions. Then the function
\begin{eqnarray}\label{p5}
W&=&\sum_{m=0}^{n-2}({\prod_{s=0}^{m-1}(n-s)\over
(-z+g(t))^{n-m}}\sum_{s=0}^m\gamma_{n-s}(t){(\int\frac{\varphi}{-z+g}\md
z)^{m-s}\over (m-s)!})x^{n-m}\nonumber\\&&
 -{g_t\over
3}(-z+g)^{-2}x^2+{(y+h(t))x\over -z+g}\nonumber\\&&
 +{2x\over
-z+g}(\int(\frac{\varphi_t}{3}+\varphi((\prod_{s=0}^{n-3}(n-s))(-z+g)^{-2}
\sum_{s=0}^{n-2}\gamma_{n-s}\nonumber\\&&\times
{(\int\frac{\varphi}{-z+g}\md z)^{n-2-s}\over
(n-2-s)!}-\frac{g_tz\varphi}{3}(-z+g)^{-2}))(-z+g)\md
z)\nonumber\\&& +y\varphi(z,t)+f(z,t)
\end{eqnarray} is a solution of the Jimbo-Miwa equation \eqref{j-m}.
\end{thm}

Suppose
\begin{equation}\label{2.35}
W=A(x,z,t)y+B(x,z,t).
\end{equation}
Then
\begin{eqnarray}\label{2.36}
W_x=A_xy+B_x,\ \ W_y=A,\ \ W_{xx}=A_{xx}y+B_{xx},\ \ W_{xy}=A_x,
\end{eqnarray}
\begin{eqnarray}\label{2.37}
W_{xz}=A_{xz}y+B_{xz}, \ W_{yt}=A_t, \ W_{xxx}=A_{xxx}y+B_{xxx}, \
W_{xxxy}=A_{xxx}.
\end{eqnarray}
Substituting \eqref{2.35}-\eqref{2.37} into \eqref{j-m}, we get
\begin{subequations}
\begin{align}
A_x^2+AA_{xx}-A_{xz}=0,\label{2.38a}\\
A_{xxx}+3A_xB_x+3AB_{xx}+2A_t-3B_{xz}=0.\label{2.38b}
\end{align}
\end{subequations}
Note that \eqref{2.38a} is the $x$-derivative inviscid Burgers
equation \cite{H-S}. A solution is
\begin{equation}\label{2.39}
A=-\frac{x-c(t)}{z-d(t)},
\end{equation}
where $c(t)$ and $d(t)$ are arbitrary functions.

Substituting \eqref{2.39} into \eqref{2.38b}, we obtain
\begin{thm} The function
\begin{eqnarray}\label{2.40}
W=-\frac{x-c(t)}{z-d(t)}y+e^{(\frac{x-c(t)}{z-d(t)})}\varphi(t)-{1\over3}d'(t)
\frac{(x-c(t))^2}{z-d(t)}+{2\over3}c'(t)x+f(t,z)\nonumber\\
\end{eqnarray} is a solution of the Jimbo-Miwa equation \eqref{j-m} for arbitrary functions $c(t),\ d(t),$ $\varphi(t)$
and $f(t,z)$.
\end{thm}

Assume
\begin{equation}\label{2.41}
W=A(x,z,t)y^2+B(x,z,t)y+D(x,z,t).
\end{equation}
Then
\begin{eqnarray}\label{2.42}
W_x=A_xy^2+B_xy+D_x,\ \ W_y=2Ay+B,
\end{eqnarray}
\begin{eqnarray}\label{2.43}
W_{xx}=2A_{xx}y^2+B_{xx}y+D_{xx},\ \ W_{xy}=2A_xy+B_x,
\end{eqnarray}
\begin{eqnarray}\label{2.44}
W_{xz}=A_{xz}y^2+B_{xz}y+D_{xz},\ \ W_{yt}=2A_ty+B_t,
\end{eqnarray}
and
\begin{eqnarray}\label{2.45}
W_{xxx}=A_{xxx}y^2+B_{xxx}y+D_{xxx},\ \ W_{xxxy}=2A_{xxx}y+B_{xxx}.
\end{eqnarray}
Substituting \eqref{2.41}-\eqref{2.45} into \eqref{j-m}, we get
\begin{subequations}
\begin{align}
A_x^2+AA_{xx}=0,\label{2.46a}\\
3A_xB_x+2AB_{xx}+A_{xx}B-A_{xz}=0,\label{2.46b}\\
2A_{xxx}+3(2A_xD_x+B_x^2)+3(2AD_{xx}+BB_{xx})+4A_t-3B_{xz}=0,\label{2.46c}\\
B_{xxx}+3B_xD_{x}+3BD_{xx}+2B_t-3D_{xz}=0.\label{2.46d}
\end{align}
\end{subequations}
Observe that
\begin{equation}\label{2.47}
A=(bx+c)^{1\over2}
\end{equation}
is a solution of \eqref{2.46a} for arbitrary functions $b=b(z,t)$
and $c=c(z,t)$. Suppose
\begin{equation}\label{2.48}
B=\sum_{n\in\mathbb{Z}}a_n(z,t)(bx+c)^n.
\end{equation}
By \eqref{2.46b}, we have
\begin{equation}\label{2.49}
B=\frac{b_z}{5b^2}(bx+c)+\frac{bc_z-b_zc}{b^2}.
\end{equation}
Denote $f=D_x$. Then \eqref{2.46c} and \eqref{2.46d} become
\begin{subequations}
\begin{align}
2A_{xxx}+6(Af)_x+3(BB_x)_x+4A_t-3B_{xz}=0,\label{2.50a}\\
3(Bf)_x+2B_t-3f_z=0.\label{2.50b}
\end{align}
\end{subequations}
Let
\begin{equation}\label{2.51}
\xi=bx+c. \end{equation} We have
\begin{equation}\label{2.52}
  A=\xi^{1\over2}
  \end{equation} by \eqref{2.47}, and
\begin{eqnarray}\label{2.53}
\xi_x=b,\ \ \xi_z=\frac{b_z}{b}\xi+b({c\over b})_z,\ \
\xi_t=\frac{b_t}{b}\xi+b({c\over b})_t.
\end{eqnarray}
Note that
\begin{equation}\label{2.54}
B=\frac{b_z}{5b^2}\xi+({c\over b})_z,\ \ B_x=\frac{b_z}{5b},\ \
B_{xz}={1\over5}(\frac{b_z}{b})_z,
\end{equation}
and
\begin{eqnarray}\label{2.55}
A_t&=&{1\over2}\xi^{-{1\over2}}(\frac{b_t}{b}\xi+b({c\over
b})_t)\nonumber\\
&=&\frac{b_t}{2b}\xi^{1\over2}+{b\over2}({c\over
b})_t\xi^{-{1\over2}}.
\end{eqnarray}
Hence, by \eqref{2.50a},
\begin{eqnarray}\label{2.56}
f=-{1\over6}({4\over3}\frac{b_t}{b^2}\xi+{1\over b}(-{3\over
5}(\frac{b_z}{b})_z+{3\over25}\frac{b_z^2}{b^2})\xi^{1\over2}+4({c\over
b})_t-{b^2\over2}\xi^{-2})+g\xi^{-{1\over2}},
\end{eqnarray}
where $g=g(z,t)$. Substituting \eqref{2.56} into \eqref{2.50b} and
checking the coefficients of $\xi^{-2}$, we get
\begin{equation}\label{2.57} b_z=0.
\end{equation}
Then
\begin{equation}\label{2.58}
f=-{1\over6}({4\over3}\frac{b_t}{b^2}\xi+4({c\over
b})_t-{b^2\over2}\xi^{-2}-6g\xi^{-{1\over2}}),
\end{equation}
 and
 \begin{equation}\label{2.59}
 B=\frac{c_z}{b}.
 \end{equation}
 Comparing the coefficients of the polynomials with respect to $\xi$ in the two sides
 of \eqref{2.50b}, we get
 \begin{subequations}
 \begin{align}
 c_{zt}b&=b_tc_z,\label{2.60a}\\
 g_z&=0.\label{2.60b}
 \end{align}
 \end{subequations}
Moreover,
\begin{equation}\label{2.61}
c=h(z)b(t)+\eta(t),\ \ g=g(t),
\end{equation}
and
\begin{eqnarray}\label{2.62}
D=-{1\over6}(\frac{b_t}{3b^3}\xi^2+2(\frac{\eta}{b})_t\frac{\xi}{b}-3\frac{g}{b}\xi^{1\over2}+{b\over2}\xi^{-1})+l(z,t).
\end{eqnarray}
\begin{thm}  For arbitrary functions $b(t)$, $h(z)$,
$\eta(t)$, $g(t)$ and $l(z,t)$, the function
\begin{eqnarray}\label{p6}
W&=&(b(t)x+h(z)b(t)+\eta(t))^{1\over2}y^2+h_zy\nonumber\\&&
-{1\over6}(\frac{b_t}{3b^3}(b(t)x+h(z)b(t)+\eta(t))^2+2(\frac{\eta(t)}{b(t)})_t\frac{b(t)x+h(z)b(t)+\eta(t)}{b(t)}\nonumber\\
&&-3\frac{g(t)}{b(t)}(b(t)x+h(z)b(t)+\eta(t))^{1\over2}+{b(t)\over2}(b(t)x+h(z)b(t)+\eta(t))^{-1})\nonumber\\&&+l(z,t)
\end{eqnarray} is a solution of the Jimbo-Miwa equation \eqref{j-m}.
\end{thm}

Let
\begin{equation}\label{2.68}
W=A(x,z,t)y^n+B(x,z,t)y+C(x,z,t),
\end{equation}
where $n\geq3$. Then
\begin{equation}\label{2.69}
W_x=A_xy^n+B_xy+C_x,\ \ W_y=nAy^{n-1}+B,
\end{equation}
\begin{equation}\label{2.70}
W_{xx}=A_{xx}y^n+B_{xx}y+C_{xx},\ \ W_{xy}=nA_xy^{n-1}+B_x,
\end{equation}
\begin{equation}\label{2.71}
W_{xz}=A_{xz}y^n+B_{xz}y+C_{xz},\ \ W_{yt}=nA_ty^{n-1}+B_t,
\end{equation} and
\begin{equation}\label{2.72}
W_{xxxy}=nA_{xxx}y^{n-1}+B_{xxx}.
\end{equation}
Substituting \eqref{2.68}-\eqref{2.72} into \eqref{j-m}, we get
\begin{eqnarray}\label{2.73}
&&nA_{xxx}y^{n-1}+B_{xxx}+3n(A_x^2+AA_{xx})y^{2n-1}\nonumber\\
&+&3((n+1)A_xB_x+nAB_{xx}+A_{xx}B)y^n
+3n(A_xC_x+AC_{xx})y^{n-1}\nonumber\\
&+&3(B_x^2+BB_{xx})y+3(B_xC_x+BC_{xx})+2nA_ty^{n-1}+2B_t\nonumber\\
&-&3A_{xz}y^n-3B_{xz}y-3C_{xz}=0.
\end{eqnarray} i.e.
\begin{subequations}
\begin{align}
A_x^2+AA_{xx}=0,\label{2.74a}\\
nAB_{xx}+(n+1)A_xB_x+A_{xx}B-A_{xz}=0,\label{2.74b}\\
A_{xxx}+3(AC_x)_x+2A_t=0,\label{2.74c}\\
B_x^2+BB_{xx}-B_{xz}=0,\label{2.74d}\\
B_{xxx}+3(BC_x)_x+2B_t-3C_{xz}=0.\label{2.74e}
\end{align}
\end{subequations}
Hence
\begin{equation}\label{2.75}
A=(\phi(z,t) x+\psi(z,t))^{1\over2}:=\xi^{1\over2}.
\end{equation}
Note that
\begin{equation}\label{2.76}
A_x={1\over2}\phi\xi^{-{1\over2}},\ \
A_{xx}=-\frac{\phi^2}{4}\xi^{-{3\over2}},\ \
A_{xxx}={3\over8}\phi^3\xi^{-{5\over2}},
\end{equation}
\begin{equation}\label{2.77}
A_t={1\over2}\frac{\phi_t}{\phi}\xi^{1\over2}+{1\over2}\frac{\phi\psi_t-\phi_t\psi}{\phi}\xi^{-{1\over2}},
\end{equation} and
\begin{equation}\label{2.78}
A_{xz}={1\over4}\phi_z\xi^{-{1\over2}}-{1\over4}(\phi\psi_z-\phi_z\psi)\xi^{-{3\over2}}.
\end{equation}
Set
\begin{equation}\label{2.79}
B=\sum_{m\in\mathbb{Z}}a_m\xi^m.
\end{equation}
Then
\begin{equation}\label{2.80}
B_x=\sum_{m\in\mathbb{Z}}m\phi a_m\xi^{m-1},\ \
B_{xx}=\sum_{m\in\mathbb{Z}}m(m-1)\phi^2a_m\xi^{m-2}.
\end{equation}
Thus
\begin{eqnarray}\label{2.81}
\sum_{m\in\mathbb{Z}}(nm(m-1)+{(n+1)\over2}m-{1\over4})a_m\phi^2\xi^m={1\over4}(\phi_z\psi-\phi\psi_z),
\end{eqnarray} i.e.
\begin{subequations}
\begin{align}
{2n+1\over4}a_1\phi^2={1\over4}\phi_z,\label{2.82a}\\
-{1\over4}a_0\phi^2=-{1\over4}(\phi\psi_z-\phi_z\psi),\label{2.82b}
\end{align}
\end{subequations} and
\begin{equation}\label{2.83}
a_m=0,\ \ \ \mbox{if}\ \ m\neq\ 0\ \mbox{or}\ 1.
\end{equation}
It deduces to
\begin{equation}\label{2.84}
B=\frac{\phi_z}{(2n+1)\phi^2}\xi+(\frac{\psi}{\phi})_z.
\end{equation}
By \eqref{2.74c}, we get
\begin{eqnarray}\label{2.85}
(AC_x)_x&=&-{1\over3}(A_{xxx}+2A_t)\nonumber\\
&=&-{1\over3}(\frac{\phi_t}{\phi}\xi^{1\over2}+\frac{\phi\psi_t-\phi_t\psi}{\phi}\xi^{-{1\over2}}+{3\over8}\phi^3\xi^{-{5\over2}}).
\end{eqnarray}
Thus
\begin{equation}\label{2.86}
C_x=-{1\over3}({2\over3}\frac{\phi_t}{\phi}\xi+2(\frac{\psi}{\phi})_t+f\xi^{-{1\over2}}-{1\over4}\phi^2\xi^{-2}).
\end{equation}
Moreover,
\begin{eqnarray}\label{2.87}
3BC_x&=&-(\frac{\phi_z\phi_t}{(2n+1)\phi^4}\xi^2+(2({\psi\over\phi})_t\frac{\phi_z}{(2n+1)\phi^2}+
{2\over3}({\psi\over\phi})_z\frac{\phi_t}{\phi^2})\xi+\frac{f\phi_z}{(2n+1)\phi^2}\xi^{1\over2}\nonumber\\
&&+2({\psi\over\phi})_z{\psi\over\phi})_t+f({\psi\over\phi})_z\xi^{1\over2}-\frac{1}{4(2n+1)}\phi_z\xi^{-1}-{1\over4}\phi^2({\psi\over\phi})_z\xi^{-2}),
\end{eqnarray}
\begin{eqnarray}\label{2.88}
3(BC_x)_x&=&-(\frac{2\phi_z\phi_t}{(2n+1)\phi^3}\xi+\frac{1}{\phi}(\frac{2}{2n+1}\phi_z(\frac{\psi}{\phi})_t+{2\over3}(\frac{\psi}{\phi})_z\phi_t)
\nonumber\\&&+{1\over2}\frac{f\phi_z}{(2n+1)\phi}\xi^{-{1\over2}}
-{1\over2}f(\frac{\psi}{\phi})_z\phi\xi^{-{3\over2}}+\frac{\phi_z\phi}{4(2n+1)}\xi^{-2}\nonumber\\
&&+{1\over2}\phi^3(\frac{\psi}{\phi})_z\xi^{-3}),
\end{eqnarray}
\begin{eqnarray}\label{2.89}
2B_t&=&\frac{2}{2n+1}(\frac{\phi_z}{\phi^2})_t\xi+\frac{2}{2n+1}\frac{\phi_z}{\phi^2}\xi_t+2(\frac{\psi}{\phi})_{zt}\nonumber\\
&=&\frac{2}{2n+1}\frac{\phi_{zt}\phi-\phi_z\phi_t}{\phi^3}\xi+\frac{2}{2n+1}\frac{\phi_z(\phi\psi_t-\phi_t\psi)}{\phi^3}+2(\frac{\psi}{\phi})_{zt},
\end{eqnarray}
\begin{eqnarray}\label{2.90}
-3C_{xz}&=&{2\over3}((\frac{\phi_t}{\phi^2})_z\xi+\frac{\phi_t}{\phi^2}(\frac{\phi_z}{\phi}\xi+\frac{\phi\psi_z-\phi_z\psi}{\phi}))
+2(\frac{\psi}{\phi})_{zt}+f_z\xi^{-{1\over2}}\nonumber\\
&&-{1\over2}f\xi^{-{3\over2}}(\frac{\phi_z}{\phi}\xi+\frac{\phi\psi_z-\phi_z\psi}{\phi})
\nonumber\\&&+{1\over2}(-\phi\phi_z\xi^{-2}+\phi^2\xi^{-3}(\frac{\phi_z}{\phi}\xi+\frac{\phi\psi_z-\phi_z\psi}{\phi}))\nonumber\\
&=&{2\over3}\frac{\phi_{tz}\phi-\phi_t\phi_z}{\phi^3}\xi+{2\over3}\frac{\phi_t}{\phi^3}(\phi\psi_z-\phi_z\psi)+2({\psi\over\phi})_{zt}
+(f_z-{1\over2}\frac{f\phi_z}{\phi})\xi^{-{1\over2}}\nonumber\\
&&-{1\over2}\frac{f(\phi\psi_z-\phi_z\psi)}{\phi}\xi^{-{3\over2}}+{1\over2}\phi(\phi\psi_z-\phi_z\psi)\xi^{-3}.
\end{eqnarray}
Hence
\begin{equation}\label{2.91}
\phi_z=0.
\end{equation}
Furthermore,
\begin{equation}\label{2.92}
3(BC_x)_x=-{2\over3}\frac{\psi_z\phi_t}{\phi^2}+{1\over2}f\psi_z\xi^{-{3\over2}}-{1\over2}\phi^2\psi_z\xi^{-3},
\end{equation}
\begin{equation}\label{2.93}
2B_t=2(\frac{\psi_z}{\phi})_t,
\end{equation} and
\begin{equation}\label{2.94}
-3C_{xz}={2\over3}\frac{\phi_t\psi_z}{\phi^2}+2(\frac{\psi_z}{\phi})_t+f_z\xi^{-{1\over2}}-{1\over2}\phi^2f\psi\xi^{-{3\over2}}.
\end{equation}
Thus
\begin{equation}\label{2.95}
(\frac{\psi_z}{\phi})_t=0,\ \ \ \ f_z=0.
\end{equation}
We get
\begin{equation}\label{2.96}
\psi=\phi(t)h(z)+g(t),\ \ \mbox{and}\ \ f=f(t).
\end{equation}
So
\begin{equation}\label{2.97}
A=(\phi x+\phi(t)h(z)+g)^{1\over2},\ \ \mbox{and}\ \ B=h_z.
\end{equation}
\begin{equation}\label{2.98}
C=-{1\over6}(\frac{\phi_t}{3\phi^3}\xi^2+2(\frac{g}{\phi})_t\frac{\xi}{\phi}-3\frac{f}{\phi}\xi^{1\over2}+{\phi\over2}\xi^{-1})+\eta(z,t).
\end{equation}

Together with Theorem 2.5, we get that
\begin{thm}
Let $\phi(t)$, $h(z)$, $g(t)$, $f(t)$ and $\eta(z,t)$ are arbitrary
functions, and let $n\geq2$ be an integer. Then the function
\begin{eqnarray}
W&=&(\phi x+\phi
h+g)^{1\over2}y^n+h_zy\nonumber-{1\over6}({1\over3}\frac{\phi_t}{\phi^3}(\phi
x+\phi h+g)^2\\&&+2(\frac{g}{\phi})_t\frac{\phi x+\phi h+g}{\phi}
-3\frac{f}{\phi}(\phi x+\phi
h+g)^{1\over2}\nonumber\\&&+{1\over2}\phi(\phi x+\phi
h+g)^{-1})+\eta(z,t)
\end{eqnarray}is a solution of the Jimbo-Miwa equation \eqref{j-m}.
\end{thm}

\subsection{Logarithmic Stable-Range Approach}

Suppose
\begin{equation}\label{2.99}
W=a(\log f)_x=a\frac{f_x}{f},
\end{equation}
for some constant $a$ and some function $f$ in $t,x,y,z$. Then
\begin{eqnarray}\label{2.100}
W_x=a\frac{ff_{xx}-f_x^2}{f^2},\ \
W_{xx}=a\frac{f^2f_{xxx}-3ff_xf_{xx}+2f_x^3}{f^3},
\end{eqnarray}
\begin{eqnarray}\label{2.101}
W_{xxx}=a\frac{f^3f_{xxxx}-4f^2f_xf_{xxx}-3f^2f_{xx}^2+12ff_x^2f_{xx}-6f_x^4}{f^4},
\end{eqnarray}
\begin{eqnarray}\label{2.102}
W_{xxxxy}&=&\frac{a}{f^5}(f^4f_{xxxxy}-f^3(f_{xxxx}f_y+4f_{xxx}f_{xy}+6f_{xx}f_{xxy}+4f_xf_{xxxy})\nonumber\\
&&+f^2(8f_xf_{xxx}f_y+6f_{xx}^2f_y+24f_xf_{xx}f_{xy}+12f_x^2f_{xxy})\nonumber\\
&&-f(36f_x^2f_{xx}f_y+24f_x^3f_{xy})+24f_x^4f_y),
\end{eqnarray}
\begin{equation}\label{2.103}
W_{xy}=\frac{a}{f^3}(f^2f_{xxy}-f(f_{xx}f_y+2f_xf_{xy})+2f_x^2f_y),
\end{equation}
\begin{equation}\label{2.104}
W_{xz}=\frac{a}{f^3}(f^2f_{xxz}-f(f_{xx}f_z+2f_xf_{xz})+2f_x^2f_z),
\end{equation}
\begin{equation}\label{2.105}
W_y=\frac{a}{f^2}(ff_{xy}-f_xf_y),
\end{equation} and
\begin{equation}\label{2.106}
W_{yt}=\frac{a}{f^3}(f^2f_{xyt}-f(f_{xy}f_t+f_xf_{yt}+f_yf_{xt})+2f_xf_yf_t).
\end{equation}
Substituting \eqref{2.99}-\eqref{2.106} into \eqref{j-m}, we find
\begin{eqnarray}\label{2.107}
&&f_{xxxxy}f^4-(f_{xxxx}f_y+4f_{xxx}f_{xy}+6f_{xx}f_{xxy}+4f_xf_{xxxy})f^3\nonumber\\
&+&(8f_xf_{xxx}f_y+6f_{xx}^2f_y+24f_xf_{xx}f_{xy}+12f_x^2f_{xxy})f^2\nonumber\\
&-&(36f_x^2f_{xx}f_y+24f_x^3f_{xy})f+24f_x^4f_y\nonumber\\
&+&3a(f^2f_{xxy}-f(f_{xx}f_y+2f_xf_{xy})+2f_x^2f_y)(ff_{xx}-f_x^2)\nonumber\\
&+&3a(ff_{xy}-f_xf_y)(f^2f_{xxx}-3ff_xf_{xx}+2f_x^3)\nonumber\\
&+&2f^2(f^2f_{xyt}-f(f_{xy}f_t+f_xf_{yt}+f_yf_{xt})+2f_xf_yf_t)\nonumber\\
&-&3f^2(f^2f_{xxz}-f(f_{xx}f_z+2f_xf_{xz})+2f_x^2f_z)=0.
\end{eqnarray} Since the left side of \eqref{2.107} is a polynomial in $f$, we set the
coefficients to be 0 and get
\begin{equation}\label{2.108}
a=2,
\end{equation}
and
\begin{subequations}
\begin{align}
f_{xxxxy}+2f_{xyt}-3f_{xxz}=0,\label{2.109a}\\
-f_{xxxx}f_y-4f_xf_{xxxy}+2f_{xxx}f_{xy}-2f_{xy}f_t\;\;\;\;\;\;\;\;\nonumber\\
-2f_xf_{xt}+3f_{xx}f_z+6f_xf_{xz}=0,\label{2.109b}\\
2f_xf_{xxx}f_y+6f_x^2f_{xxy}-6f_xf_{xx}f_{xy}+4f_xf_yf_t-6f_x^2f_z=0.\label{2.109c}
\end{align}
\end{subequations}
Simplifying \eqref{2.109a}-\eqref{2.109c}, we have
\begin{subequations}
\begin{align}
f_{xxxy}+2f_{yt}-3f_{xz}=0,\label{2.110a}\\
f_{xxx}f_y-3f_{xx}f_{xy}+3f_xf_{xxy}+2f_yf_t-3f_xf_z=0.\label{2.110b}
\end{align}
\end{subequations}

Let
\begin{equation}\label{4.1}
f=\sum_{m=0}^nA_m(y,z,t)x^m,
\end{equation}
where
\begin{equation}\label{4.2}
A_n=1
\end{equation} and
\begin{equation}\label{4.3}
A_m(y,z,t)=A_m(t)
\end{equation} for $m=1,\ldots,n-1.$
We set $A_{n+s}=0$ for $s>0$. Then
\begin{eqnarray}\label{4.4}
f_x=\sum_{m=0}^n(m+1)A_{m+1}x^m,\ \ \ \ f_y=A_{0y},
\end{eqnarray}
\begin{eqnarray}\label{4.5}
f_t=\sum_{m=0}^nA_{mt}x^m,\ \ \ \ f_z=A_{0z},
\end{eqnarray}
\begin{equation}\label{4.6}
f_{xx}=\sum_{m=0}^n(m+2)(m+1)A_{m+2}x^m,
\end{equation}
\begin{equation}\label{4.7}
f_{xy}=f_{xz}=0,\ \ \ \ f_{yt}=A_{0yt},
\end{equation} and
\begin{equation}\label{4.8}
f_{xxx}=\sum_{m=0}^n(m+3)(m+2)(m+1)A_{m+3}x^m.
\end{equation}
Substituting \eqref{4.4}-\eqref{4.8} into \eqref{2.110a} and
\eqref{2.110b}, we get that
\begin{equation}\label{4.9}
A_{0yt}=0,
\end{equation} and
\begin{equation}\label{4.10}
A_{0y}((m+3)(m+2)(m+1)A_{m+3}+2A_{mt})=3(m+1)A_{0z}A_{m+1}.
\end{equation}
Thus we can assume that
\begin{equation}\label{4.11}
{A_{0z}\over A_{0y}}=k,
\end{equation}
where $k$ is an constant.

By induction, we get that
\begin{equation}\label{4.12}
A_{n-s}=\sum_{r=0}^s\sum_{p=0}^{\llbracket{r\over2}\rrbracket}
(\prod_{l=0}^{s-1}(n-l))(-1)^p{s-r\choose
p}({1\over2})^p({3\over2}k)^{s-r-p}k_{n-r+2p}{t^{s-r}\over (s-r)!}
\end{equation}
for $s=0,1,\ldots,n-1.$ Here $k_n=1$ and $k_1,\ldots,k_{n-1}$ are
arbitrary constants. Moreover,
\begin{eqnarray}\label{4.13}
A_0&=&\eta(y+kz)\nonumber\\&&+
\sum_{r=0}^n\sum_{p=0}^{\llbracket{r\over2}\rrbracket}
(\prod_{l=0}^{n-1}(n-l))(-1)^p{n-r\choose
p}({1\over2})^p({3\over2}k)^{n-r-p}k_{n-r+2p}{t^{n-r}\over
(n-r)!},\nonumber\\
\end{eqnarray} where $\eta(y+kz)$ is an arbitrary function of $y+kz$, and $k_0$ is an arbitrary constant.

In particular, we set
\begin{equation}\label{2.111}
f=x+B(y,z,t).
\end{equation}
By \eqref{2.110a} and \eqref{2.110b},
\begin{subequations}
\begin{align}
B_{yt}&=0,\label{2.112a}\\
2B_yB_t&=3B_z.\label{2.112b}
\end{align}
\end{subequations}
So, we have that
\begin{equation}\label{2.113}
B=g(y,z)+h(t,z),
\end{equation}
and
\begin{equation}\label{2.114}
2g_yh_t=3(g_z+h_z).
\end{equation}

Assume that $g$ is a polynomial in variable $y$. If
\begin{equation}\label{2.115}
g=C(z)y+D(z),
\end{equation}
then by \eqref{2.114},
\begin{equation}\label{2.116}
2Ch_t=3D_z+3h_z,
\end{equation}
and $C$ is a constant.  Differentiating \eqref{2.116} with respect
to $t$, we obtain
\begin{equation}\label{2.117}
{2\over3}C=\frac{(h_t)_z}{(h_t)_t}.
\end{equation}
Thus
\begin{subequations}
\begin{align}
h_t&=\phi(t+{2\over3}Cz),\label{2.118a}\\
h_z&={2\over3}c\phi(t+{2\over3}Cz)+\psi'(z),\label{2.118b}
\end{align}
\end{subequations}
where $\phi$ and $\psi$ are arbitrary functions. Since
\begin{equation}\label{2.119}
2C\phi(t+{2\over3}Cz)-2C\phi(t+{2\over3}Cz)-3\psi'(z)=3D_z,
\end{equation}
we have that
\begin{equation}\label{2.120}
g=Cy-\psi(z)+k,\ \ h=\rho(t+{2\over3}Cz)+\psi(z).
\end{equation}
where $\rho$ and $\psi$ are arbitrary functions.

If
\begin{equation}\label{2.121}
g=\sum_{m=0}^na_m(z)y^m,\ \ \ \ \ \ (n\geq2)
\end{equation}
then  by \eqref{2.112b}, we have
\begin{equation}\label{2.122}
a_{n-m}=\sum_{r=0}^m(\prod_{s=0}^{m-1}(n-s))(\frac{2b}{3})^mk_r\frac{z^{m-r}}{(m-r)!}-\delta_{n,m}F(z)
\end{equation}
for $m=0,1,\ldots,n-1$, and
\begin{equation}\label{2.123}
h=bt+F(z),
\end{equation} where $F(z)$ is an arbitrary function.

Take
\begin{equation}\label{2.124}
f=Ay+B.
\end{equation}
According to \eqref{2.110a} and \eqref{2.110b},
\begin{subequations}
\begin{align}
A_xA_z=0,\label{2.125a}\\
A_{xxx}+2A_t-3B_{xz}=0,\label{2.125b}\\
AA_{xxx}+2AA_t-3(A_xB_z+A_zB_x)=0,\label{2.125c}\\
AB_{xxx}-3A_xB_{xx}+3A_{xx}B_x+2AB_t-3B_xB_z=0.\label{2.125d}
\end{align}
\end{subequations}
If $A_x=0$, the solution will be the same as the preceding case.
Thus we suppose
\begin{equation}\label{2.126}
A_z=0.
\end{equation}
Moreover, we assume
\begin{equation}\label{2.127}
A=e^{ax+bt}.
\end{equation} Then by \eqref{2.125b} and \eqref{2.125c}, we obtain
\begin{equation}\label{2.128}
B_{xz}=\frac{a^3+2b}{3}e^{ax+bt}\ \ \ \ \mbox{and}\ \ \
B_z=\frac{a^3+2b}{3a}e^{ax+bt}.
\end{equation} Thus
\begin{equation}\label{2.129}
B=\frac{a^3+2b}{3a}Az+\phi(t,x).
\end{equation}
Substituting \eqref{2.129} into \eqref{2.125d}, we get
\begin{eqnarray}\label{2.130}
\phi_{xxx}-3a\phi_{xx}+3a^2\phi_x+2\phi_t-\frac{a^3+2b}{a}\phi_x=0.
\end{eqnarray}
It is a flag type equation \cite{X2}. We can get a basis of its
polynomial solution space as follows
\begin{eqnarray}\label{2.131}
\phi(t,x)&=&\sum_{r_1,r_2,r_3=0}^\infty(-1)^{r_1+r_3}\frac{3^{r_2}a^{r_2}(a^3-b)^{r_3}}{2^{r_1+r_2}}
\frac{\prod\limits_{s=0}^{3r_1+2r_2+r_3+1}(n-s)}{(r_1+r_2+r_3)!}\nonumber\\&&\times
x^{n-3r_1-2r_2-r_3}t^{r_1+r_2+r_3}.
\end{eqnarray}
We write the results in this subsection as follows
\begin{thm}
The functions
\begin{eqnarray}\label{p15}
W_1&=&2(\sum_{s=1}^n\sum_{r=0}^s\sum_{p=0}^{\llbracket{r\over2}\rrbracket}
(\prod_{l=0}^{s}(n-l))(-1)^p{s-r\choose
p}({1\over2})^p({3\over2}k)^{s-r-p}k_{n-r+2p}\nonumber\\&&
\times{t^{s-r}\over
(s-r)!}x^{n-s})(\sum_{s=0}^n\sum_{r=0}^s\sum_{p=0}^{\llbracket{r\over2}\rrbracket}
(\prod_{l=0}^{s-1}(n-l))(-1)^p{s-r\choose
p}({1\over2})^p({3\over2}k)^{s-r-p}\nonumber\\&& \times
k_{n-r+2p}{t^{s-r}\over (s-r)!}x^{n-s}+\eta(y+kz))^{-1}
\end{eqnarray}
\begin{eqnarray}\label{p8}
W_2=2(x+Cy+k+\rho(t+{2\over3}Cz))^{-1},
\end{eqnarray}
\begin{eqnarray}\label{p9}
W_3=2(x+\sum_{m=0}^n\sum_{r=0}^m(\prod_{s=0}^{m-1}(n-s))({2b\over
3})^mk_r{z^{m-r}\over(m-r)!}y^{n-m}+bt)^{-1}
\end{eqnarray} and
\begin{eqnarray}\label{p10}
W_4=2{ae^{ax+bt}y+{a^3+2b\over3}e^{ax+bt}z+\phi_x(t,x)\over
e^{ax+bt}y+{a^3+2b\over3a}e^{ax+bt}z+\phi(t,x)}
\end{eqnarray} are solutions of \eqref{j-m}, where
$\rho(t+{2\over3}Cz)$ is an arbitrary function of $t+{2\over3}Cz$,
$\eta(y+kz)$ is an arbitrary function of $y+kz$, the numbers $C$,
$k$, $k_r$, $a$ and $b$ are constants, and the function $\phi$ is
given by \eqref{2.131}.
\end{thm}

\section{Konopelchenko-Dubrovsky Equations}
\subsection{Stable-Range Approach}

By (1.2b), we take the potential form
\begin{equation}\label{3.1}
u=W_x,\ \ \ \ v=W_y.
\end{equation}
Then the Konopelchenko-Dubrovsky equations \eqref{k-da} and
\eqref{k-db} are equivalent to
\begin{equation}\label{3.2}
W_{xt}-W_{xxxx}-6bW_xW_{xx}+{3\over2}a^2W_x^2W_{xx}-3W_{yy}+3aW_{xx}W_y=0.
\end{equation}

Suppose
\begin{equation}\label{3.4}
W=Ax^2+Bx+C
\end{equation}
for some functions $A,\;B$ and $C$ in $t$ and $y$. Note that
\begin{equation}\label{3.5}
W_x=2Ax+B,\ \ W_{xx}=2A,\ \ W_y=A_yx^2+B_yx+C_y,
\end{equation}
\begin{equation}\label{3.6}
W_{yy}=A_{yy}x^2+B_{yy}x+C_{yy},\ \ W_{xt}=2A_tx+B_t.
\end{equation}
Substituting \eqref{3.4}-\eqref{3.6}, we find
\begin{eqnarray}\label{3.7}
&&2A_tx+B_t-12Ab(2Ax+B)+3a^2A(2Ax+B)^2\nonumber\\&-&3(A_{yy}x^2+B_{yy}x+C_{yy})
+6aA(A_yx^2+B_yx+C_y)=0.
\end{eqnarray}
Hence
\begin{subequations}
\begin{align}
4a^2A^3-A_{yy}+2aAA_y=0,\label{3.8a}\\
2A_t-24A^2b+12a^2A^2B-3B_{yy}+6aAB_y=0,\label{3.8b}\\
B_t-12ABb+3a^2AB^2-3C_{yy}+6aAC_y=0,\label{3.8c}
\end{align}
\end{subequations}
Observe that
\begin{equation}\label{3.9}
A=\frac{1}{ay+\psi(t)}
\end{equation} and
\begin{equation}\label{3.10}
A=\frac{1}{-2ay+\psi(t)}
\end{equation} are two of solutions of \eqref{3.8a}, where $\psi(t)$ is an arbitrary function.
Substituting these two solutions into \eqref{3.8b}, we get that
\begin{equation}\label{3.11}
B=f_{-1}(t)(ay+\psi)^{-1}+f_0+f_4(t)(ay+\psi)^4
\end{equation}
or
\begin{equation}\label{3.12}
B=f_{-1}(t)(-2ay+\psi)^{-1}+f_0+f_1(t)(-2ay+\psi),
\end{equation}
where $f_0=(\psi_t+12b)/(6a^2)$. Thus we have
\begin{eqnarray}\label{3.13}
C&=&{f_{-1}^2\over4}(ay+\psi)^{-1}-{1\over
3a^2}(-{1\over3}f_{-1}f_4-4bf_{-1}+2a^2f_{-1}f_0)\log(ay+\psi)\nonumber\\
&&-{1\over
2a^2}({1\over3}f_{(-1)t}-4bf_0+a^2f_0^2)(ay+\psi)+{\phi(z,t)\over
3a}(ay+\psi)^3\nonumber\\&&
+{f_{-1}f_4\over2}(ay+\psi)^4+{1\over10a^2}({4\over3}f_4\psi_t-4bf_4+2a^2f_0f_4)(ay+\psi)^5\nonumber\\
&&+{f_{4t}\over54a^2}(ay+\psi)^6+{f_4^2\over54}(ay+\psi)^9+\varsigma(z,t)
\end{eqnarray} or
\begin{eqnarray}\label{3.14}
C&=&{1\over4}f_{-1}^2(-2ay+\psi)^{-1}-{1\over2a}\phi(z,t)\log(-2ay+\psi)\nonumber\\&&+
{1\over8a^2}(-{1\over3}f_{-1}\psi_t-4bf_{-1}+2a^2f_0f_{-1})\log^2(-2ay+\psi)\nonumber\\&&
+{1\over4a^2}({f_{(-1)t}\over3}-4bf_0+a^2(2f_{-1}f_1+f_0^2))(-2ay+\psi)
\nonumber\\&&+
{1\over16a^2}({1\over3}(f_{0t}+f_1\psi_t)-4bf_1+2a^2f_0f_1)(-2ay+\psi)^2\nonumber\\&&+
{1\over36a^2}({1\over3}f_{1t}+a^2f_1^2)(-2ay+\psi)^3+\varsigma(z,t),
\end{eqnarray} where $\phi(z,t)$ and $\varsigma(z,t)$ are arbitrary
functions.
\begin{thm}
The functions
\begin{eqnarray}\label{p11}
W_1&=&(ay+\psi(t))^{-1}x^2+(f_{-1}(t)(ay+\psi)^{-1}+f_0+f_4(t)(ay+\psi)^4)x\nonumber\\&-&
{f_{-1}^2\over4}(ay+\psi)^{-1}-{1\over
3a^2}(-{1\over3}f_{-1}f_4-4bf_{-1}+2a^2f_{-1}f_0)\log(ay+\psi)\nonumber\\
&&-{1\over
2a^2}({1\over3}f_{(-1)t}-4bf_0+a^2f_0^2)(ay+\psi)+{\phi(z,t)\over
3a}(ay+\psi)^3\nonumber\\&&
+{f_{-1}f_4\over2}(ay+\psi)^4+{1\over10a^2}({4\over3}f_4\psi_t-4bf_4+2a^2f_0f_4)(ay+\psi)^5\nonumber\\
&&+{f_{4t}\over54a^2}(ay+\psi)^6+{f_4^2\over54}(ay+\psi)^9+\varsigma(z,t)
\end{eqnarray} and
\begin{eqnarray}\label{p12}
W_2&=&(ay+\psi(t))^{-1}x^2+(f_{-1}(t)(-2ay+\psi)^{-1}+f_0+f_1(t)(-2ay+\psi))x\nonumber\\&&
{1\over4}f_{-1}^2(-2ay+\psi)^{-1}-{1\over2a}\phi(z,t)\log(-2ay+\psi)\nonumber\\&&+
{1\over8a^2}(-{1\over3}f_{-1}\psi_t-4bf_{-1}+2a^2f_0f_{-1})\log^2(-2ay+\psi)\nonumber\\&&
+{1\over4a^2}({f_{(-1)t}\over3}-4bf_0+a^2(2f_{-1}f_1+f_0^2))(-2ay+\psi)
\nonumber\\&&+
{1\over16a^2}({1\over3}(f_{0t}+f_1\psi_t)-4bf_1+2a^2f_0f_1)(-2ay+\psi)^2\nonumber\\&&+
{1\over36a^2}({1\over3}f_{1t}+a^2f_1^2)(-2ay+\psi)^3+\varsigma(z,t)
\end{eqnarray} are solutions of \eqref{3.2}, where
$f_0=(\psi_t+12b)/(6a^2)$. The functions $\psi(t)$, $f_{-1}(t)$,
$f_1(t)$, $f_4(t)$, $\phi(z,t)$ and $\varsigma(z,t)$ are arbitrary
functions.
\end{thm}

\subsection{Logarithmic Stable-Range Approach}

Assume
\begin{equation}\label{3.15}
W=m\log f
\end{equation} for some real number $m$ and function $f$ in $t,x$ and $y$.
Then
\begin{equation}\label{3.16}
u=W_x=m\frac{f_x}{f},\ \ v=W_y=m\frac{f_y}{f},
\end{equation}
Note that
\begin{equation}\label{3.17}
(\frac{f_x}{f})_t=\frac{f_{xt}f-f_xf_t}{f^2},\ \
(\frac{f_x}{f})_x=\frac{f_{xx}f-f_x^2}{f^2},
\end{equation}
\begin{equation}\label{3.18}
(\frac{f_x}{f})_{xx}=\frac{f^2f_{xxx}-3ff_xf_{xx}+2f_x^3}{f^3},\ \
(\frac{f_y}{y})_y=\frac{f_{yy}f-f_y^2}{f^2},
\end{equation} and
\begin{equation}\label{3.19}
(\frac{f_x}{f})_{xxx}=\frac{f^3f_{xxxx}-f^2(4f_xf_{xxx}+3f_{xx}^2)+12ff_x^2f_{xx}-6f_x^4}{f^4}.
\end{equation}
Substituting \eqref{3.15}-\eqref{3.19} into \eqref{3.2}, we find
\begin{eqnarray}\label{3.20}
&&(ff_{xt}-f_xf_t)f^2-(f_{xxxx}f^3-(4f_xf_{xxx}+3f_{xx}^2)f^2+12f_x^2f_{xx}f-6f_x^4)\nonumber\\
&-&6mbff_x(f_{xx}f-f_x^2)+{3\over2}a^2m^2f_x^2(f_{xx}f-f_x^2)-3f^2(f_{yy}f-f_y^2)\nonumber\\
&+&3amff_y(f_{xx}f-f_x^2)=0.
\end{eqnarray}
We assume that the coefficients of the polynomial with respect to
$f$ in the left side of \eqref{3.20} are 0. Then we get
\begin{equation}\label{3.21}
m=\pm{2\over a},
\end{equation}
and
\begin{subequations}
\begin{align}
f_{xt}-f_{xxxx}-3f_{yy}=0,\label{3.22a}\\
-af_xf_t+4af_xf_{xxx}+3af_{xx}^2\mp12bf_xf_{xx}+3af_y^2\pm6af_yf_{xx}=0,\label{3.22b}\\
(-af_{xx}\pm2bf_x\mp af_y)f_x^2=0.\label{3.22c}
\end{align}
\end{subequations}
Simplifying the above system, we obtain
\begin{subequations}
\begin{align}
f_{xt}-f_{xxxx}-3f_{yy}=0,\label{3.23a}\\
-af_t+4af_{xxx}\mp12bf_{xx}+\frac{12b^2}{a}f_x=0,\label{3.23b}\\
f_{xx}\pm f_y\mp{2b\over a}f_x=0.\label{3.23c}
\end{align}
\end{subequations}
The equations \eqref{3.23b} and \eqref{3.23c} imply \eqref{3.23a}.
Note
\begin{equation}\label{3.24}
f_{xxx}=\frac{4b^2}{a^2}f_x-{2b\over a}f_y\mp f_{xy}
\end{equation}
by \eqref{3.23c}. Then
\begin{eqnarray}\label{3.25}
f_{xxxx}&=&\pm{2b\over a}f_{xxx}\mp f_{xxy}\nonumber\\
&=&\pm{8b^3\over a^3}f_x\mp{4b^2\over a^2}f_y-{4b\over
a}f_{xy}+f_{yy}.
\end{eqnarray}
Moveover, by \eqref{3.23b}, \eqref{3.24} and \eqref{3.25}, we get
\begin{eqnarray}\label{3.26}
f_{xt}&=&4f_{xxxx}\mp{12b\over a}f_{xxx}+{12b^2\over
a^2}f_{xx}\nonumber\\
&=&\pm{8b^3\over a^3}f_x\mp{4b^2\over a^2}f_y-{4b\over
a}f_{xy}+4f_{yy}\nonumber\\
&=&f_{xxxx}+3f_{yy}.
\end{eqnarray}
Thus the system (3.24) can be written as
\begin{subequations}
\begin{align}
f_{xx}\pm f_y=\pm{2b\over b}f_x,\label{3.27a}\\
f_t=4f_{xxx}\mp{12b\over a}f_{xx}+{12b^2\over a^2}f_x.\label{3.27b}
\end{align}
\end{subequations}
Let \begin{equation}\label{3.27.1}
 m={2\over a}.
\end{equation}
Then the system (3.28) becomes
\begin{subequations}
\begin{align}
f_{xx}+f_y={2b\over b}f_x,\label{3.27.2a}\\
f_t=4f_{xxx}-{12b\over a}f_{xx}+{12b^2\over a^2}f_x.\label{3.27.2b}
\end{align}
\end{subequations}
Note that the case $m=-2/a$ can be translated into the case $m=2/a$
if we set $h(x,y,t)=f(-x,-y,-t)$. Thus it is sufficiently to
calculate the case $m=2/a$.
 We assume
\begin{equation}\label{3.28}
f=\sum_{m=0}^na_m(y,t)\xi^m,\qquad\xi=x+{2b\over a}y+{12b^2\over
a^2}t.
\end{equation}
 Then
\begin{eqnarray}\label{3.29}
f_x=\sum_{m=0}^{n-1}(m+1)a_{m+1}\xi^m,\ \ \
f_{xx}=\sum_{m=0}^{n-2}(m+2)(m+1)a_{m+2}\xi^m,
\end{eqnarray}
\begin{eqnarray}\label{3.30}
f_{xxx}=\sum_{m=0}^{n-3}(m+3)(m+2)(m+1)a_{m+3}\xi^m,
\end{eqnarray}
\begin{equation}\label{3.34}
f_t=\sum_{m=0}^{n-1}({12b^2\over a^2}(m+1)a_{m+1}+a_{mt})\xi^m.
\end{equation}
By \eqref{3.27.2a} and \eqref{3.27.2b}, we find
\begin{subequations}
\begin{align}
a_{my}&=-(m+2)(m+1)a_{m+2},\label{3.35a}\\
a_{mt}&=4(m+3)(m+2)(m+1)a_{m+3}-{12b\over
a}(m+2)(m+1)a_{m+2},\label{3.35b}
\end{align}
\end{subequations}
where we have supposed that $a_{n+l}=0$ for $l>0$. Hence
\begin{equation}\label{3.36}
a_n=b_n\ \ \ \ \ \ \mbox{and}\ \ \ \ a_{n-1}=b_{n-1}
\end{equation}
are constants. Let
\begin{equation} \label{3.37}
\left\{ \begin{aligned}
         \eta &= y+{12b\over a}t \\
         t&=t
\end{aligned} \right.
\end{equation}
Then we get that
\begin{subequations}
\begin{align}
a_{m\eta}&=-(m+2)(m+1)a_{m+2},\label{3.38a}\\
a_{mt}&=4(m+3)(m+2)(m+1)a_{m+3}\label{3.38b}
\end{align}
\end{subequations}by \eqref{3.35a} and \eqref{3.35b}.

Denote
\begin{equation}\label{3.39}
d(m,k)=\llbracket k/ m\rrbracket,\ \ \ \ r(m,k)=k-\llbracket
k/m\rrbracket
\end{equation}
for $m,k\in \mathbb{Z}^+$. Then by induction, we obtain
\begin{eqnarray}\label{3.40}
&&a_{n-k}(y,t)=\sum_{p=0}^{d(3,k)-1}\sum_{q=0}^{d(2,3p+r(3,k))}{(-1)^{d(2,3p+r(3,k))-q}4^{d(3,k)-p}\over
(d(3,k)-p)!(d(2,3p+r(3,k))-q)!}\nonumber\\&\times&(\prod_{l=3p+r(3,k)}^l(n-l))\nonumber\\
&\times&
(\prod_{s=2q+r(2,3p+r(3,k))}^{3p+r(3,k)-1}(n-s))b_{n-2q-r(2,3p+r(3,k))}\eta^{d(2,3p+r(3,k))-p}t^{d(3,k)-p}\nonumber\\
&+&\sum_{q=0}^{d(2,k)}{(-1)^{d(2,k)-q}\over
(d(2,k)-q)!}(\prod_{l=2q-r(2,k)}^{k-1}(n-l))b_{n-2q-r(2,k)}\eta^{d(2,k)-q},
\end{eqnarray}
where $b_{n-m}$ are constants.
\begin{thm} For any positive integer $n$, the functions
\begin{eqnarray}\label{p13}
W=\pm{2\over a}\log(\sum_{m=0}^na_m(\pm y,\pm t)(\pm(x+{2b\over
a}y+{12b^2\over a^2}t))^m)
\end{eqnarray} are solutions of \eqref{3.2}, where $a_m$ are given
by \eqref{3.40}.
\end{thm}

Next we assume
\begin{equation}\label{3.41} f=\sum_{m=0}^nA_my^m,
\end{equation}
where $A_m$ are functions in $t$ and $x$. Then by \eqref{3.27a} and
\eqref{3.27b},
\begin{subequations}
\begin{align}
A_{nxx}&={2b\over a}A_{nx},\label{3.42a}\\
A_{mxx}&={2b\over a}A_{mx}-(m+1)A_{m+1},\label{3.42b}\\
A_{mt}&={4b^2\over a^2}A_{mx}+{4b\over
a}(m+1)A_{m+1}-4(m+1)A_{(m+1)x},\label{3.42c}
\end{align}
\end{subequations}
for $m=0,\ldots,n-1$.

Write
\begin{equation}\label{3.46}
A_m=g_m(x,t)\exp({2b\over a}x+{8b^3\over a^3}t)
\end{equation}
for $m=0,\ldots,n$. Then by \eqref{3.42a}, \eqref{3.42b} and
\eqref{3.42c},
\begin{subequations}
\begin{align}
g_{mxx}&=-{2b\over a}g_{mx}-(m+1)g_{m+1},\label{3.47a}\\
g_{mt}&={12b^2\over a^2}g_{mx}+{12b\over
a}g_{mxx}+4g_{mxxx}.\label{3.47b}
\end{align}
\end{subequations}
We assume
\begin{equation}\label{3.48}
g_{n-m}(x,t)=\sum_{s=0}^mB_s^{n-m}(t)x^s
\end{equation}
for $m=0,\ldots,n$, where $B_s^{n-m}$ are functions in $t$. Thus
\begin{subequations}
\begin{align}
B_{s,t}^0&={12b^2\over a^2}(s+1)B_{s+1}^0+{12b\over
a}(s+2)(s+1)B_{s+2}^0\nonumber\\&+4(s+3)(s+2)(s+1)B_{s+3}^0,\label{3.49a}\\
B_s^{n-m+1}&=-{(s+2)(s+1)\over n-m+1}B_{s+2}^{n-m}-{2b\over
a}{s+1\over n-m+1}B_{s+1}^{n-m}\label{3.49b}
\end{align}
\end{subequations}
Firstly,
\begin{equation}\label{3.50}
g_0=\sum_{s=0}^nB_s^0x^s.
\end{equation}
Note that we can write
\begin{equation}\label{3.51}
B_{n-m}^0=(\prod_{l=0}^{m-1}(n-l))\sum_{p=0}^mc_{n-p}d_{m-p},
\end{equation}
where $c_{n-p}$ are constants, and
\begin{equation}\label{3.52}
d_0=1,\ \ \ \ d_1={12b^2\over a^2}t.
\end{equation}
Observe that
\begin{equation}\label{3.53}
d_m={12b^2\over a^2}\int d_{m-1}\md t+{12b\over a}\int d_{m-2}\md
t+4\int d_{m-3}\md t.
\end{equation}
Thus we can write
\begin{equation}\label{3.54}
d_m=\sum_{s=0}^{\llbracket{2m\over3}\rrbracket}e_{m,s}12^{m-s}({b\over
a})^{2m-3s}{t^{m-s}\over (m-s)!}.
\end{equation}
Then
\begin{subequations}
\begin{align}
e_{0,0}&=1,\label{3.55a}\\
e_{0,p}&=e_{l,-p}=e_{-l,p}=0 \ \ \ \ \mbox{for}\ \ p>0\ \mbox{and}\
l>0.\label{3.55b}
\end{align}
\end{subequations}
and
\begin{equation}\label{3.56}
e_{m,k}=e_{m-1,k}+e_{m-2,k-1}+{1\over 3}e_{m-3,k-2}.
\end{equation}
for $m>0$ and $k>0$ again by \eqref{3.49a}. Hence
\begin{equation}\label{3.57}
e_{m,k}=\sum_{s=0}^k({1\over3})^s{k-s \choose s}{m-k\choose k-s}.
\end{equation}
Thus
\begin{eqnarray}\label{3.58}
d_m=\sum_{k=0}^{\llbracket{2m\over3}\rrbracket}\sum_{s=0}^k12^{m-k}({1\over3})^s{k-s
\choose s}{m-k\choose k-s}({b\over a})^{2m-3k}{t^{m-k}\over (m-k)!}.
\end{eqnarray}
So we have
\begin{eqnarray}\label{3.59}
B_{n-m}^0&=&(\prod_{l=0}^{m-1}(n-l))\sum_{p=0}^m\sum_{k=0}^{\llbracket{2m-p\over3}\rrbracket}\sum_{s=0}^k
c_{n-p}12^{m-p-k}\nonumber\\
&\times&({1\over3})^s{k-s \choose s}{m-p-k\choose k-s}({b\over
a})^{2m-2p-3k}{t^{m-p-k}\over (m-p-k)!},
\end{eqnarray}
\begin{eqnarray}\label{3.60}
g_0&=&\sum_{m=0}^nB_{n-m}^0x^{n-m}\nonumber\\
&=&\sum_{m=0}^n(\prod_{l=0}^{m-1}(n-l))\sum_{p=0}^m\sum_{k=0}^{\llbracket{2m-p\over3}\rrbracket}\sum_{s=0}^k
c_{n-p}12^{m-p-k}\nonumber\\
&\times&({1\over3})^s{k-s \choose s}{m-p-k\choose k-s}({b\over
a})^{2m-2p-3k}{t^{m-p-k}\over (m-p-k)!}x^{n-m}.
\end{eqnarray}
Now we calculate
\begin{equation}\label{3.61}
g_{n-q}=\sum_{r=0}^qB_r^{n-q}x^r
\end{equation}
for $q=0,1,\ldots,n-1$,
\begin{equation}\label{3.62}
B_r^{n-q}=-{(r+2)(r+1)\over n-q}B_{r+2}^{n-q-1}-{2b\over a}{r+1\over
n-q}B_{r+1}^{n-q-1}
\end{equation}
for $q=0,1,\ldots,n-1$. Thus
\begin{equation}\label{3.63}
B_r^m=\sum_{s=0}^{2^m}(\prod_{l=0}^{m+s}(r+l)){{m\choose s}\over
m!}({2b\over a})^sB_{r+m+s}^0
\end{equation}
where
\begin{equation}\label{3.64}
B_{n+l}^0=0
\end{equation}
for $l>0$.
\begin{thm}The functions
\begin{eqnarray}\label{p14}
W&=&\pm{2\over a}\log[\sum_{s=0}^nB_s^0(\pm t)(\pm
x)^s\exp(\pm({2b\over a}x +{8b^3\over a^3}t))\nonumber\\&&
+\sum_{m=1}^n\sum_{r=0}^{n-m}\sum_{s=0}^{2^m}(\prod_{l=0}^{m+s}(r+l)){{m\choose
s}\over m!}({2b\over a})^s\nonumber\\&&\times B_{r+m+s}^0(\pm
t)\exp(\pm({2b\over a}x +{8b^3\over a^3}t))(\pm y)^m]
\end{eqnarray} are solutions of \eqref{3.2}, where $B_s^0$ are given
by \eqref{3.59} and \eqref{3.64}.
\end{thm}

 \textbf{Acknowledgement:} I would like to thank
Professor Xiaoping Xu for his advice and suggesting this research
topic.


\begin{thebibliography}{50}
\small

\bibitem{A}
{\sc M. A. Abdou,} Generalized solitonary and periodic solutions for
nonlinear partial differential equations by the Exp-function method,
{\it Nonlinear Dyn.} {\bf 52} (2008), 1-9.

\bibitem{D-G-R-W}
{\sc B. Dorizzi, B. Grammaticos, A. Ramani and P. Winternitz,} Are
all the equations of the Kadomtsev-Petviashvili hierarchy
integrable? {\it J. Math. Phys.} {\bf 12} (1986), 2848--2852.

\bibitem{F}
{\sc E. Fan,} An algebraic method for finding a series of exact
solutions to integrable and nonintegrable nonlinear evolution
equations, {\it J. Phys. A} {\bf 36} (2003), 7009--7026.

\bibitem{H-O}
{\sc W. Hong and K. Oh,} New solitonic solutions to a
(3+1)-dimensional Jimbo-Miwa equation, {\it Comput. Math. Appl.}
{\bf 39} (2000), 29--31.

\bibitem{H-S}
{\sc J. Hunter and R. Saxton,} Dynamics of Director Fields, {\it
SIAM J. Appl. Math.} {\bf 51} (1991), 1498--1521.

\bibitem{J-M}
{\sc M. Jimbo and T. Miwa,} Solitons and infinite dimensional Lie
algebras, {\it Publ. RIMS. Kyoto Univ.} {\bf 19} (1983), 943--1001.

\bibitem{K-D}
{\sc B. Konopelchenko and V. Dubrovsky,} Some new integrable
nonlinear evolution equations in 2+1 dimensions, {\it Phys. Lett.}
{\bf 102A} (1984), 15--17.

\bibitem{L-L-W}
{\sc J. Lin, S. Lou and K. Wang,} Multi-soliton solutions of the
Konopelchenko-Dubrovsky equation, {\it Chin. Phys. Lett.} {\bf 18}
(2001), 1173--1175.

\bibitem{L-W}
{\sc S. Lou and J. Weng,} Generalized $W_\infty$ symmetry algebra of
the conditionally integrable nonlinear evolution equation, {\it J.
Math. Phys.} {\bf 36} (1995), 3492--3497.

\bibitem{M}
{\sc A. Maccari,} A new integrable Davey-Stewartson-type equation,
{\it J. Math. Phys.} {\bf 40} (1999), 3971--3977.

\bibitem{R-W}
{\sc J. Rubin and P. Winternitz,} Point symmetries of conditionally
integrable nonlinear evolution equations, {\it J. Math. Phys.} {\bf
31} (1990), 2085--2090.

\bibitem{S-Z1}
{\sc L. Song and H. Zhang,} New exact solutions for the
Konopelchenko-Dubrovsky equation using an extended Riccati equation
rational expansion method and symbolic computation, {\it Appl. Math.
Comput.} {\bf 187} (2007), 1373--1388.

\bibitem{S-Z2}
{\sc L. Song and H. Zhang,} Application of the extended homotopy
perturbation method to a kind of nonlinear evolution equations, {\it
Appl. Math. Comput.} {\bf 197} (2008), 87-95.

\bibitem{W}
{\sc A. Wazwaz,} New kinks and solitons solutions to the
(2+1)-dimensional Konopelchenko-Dubrovsky equation, {\it Math.
Comput. Model.} {\bf 45} (2007), 473--479.

\bibitem{W-Z}
{\sc D. Wang and H. Zhang,} Further improved F-expansion method and
new exact solutions of Konopelchenko-Dubrovsky equation, {\it Chaos,
Solitons and Fractals} {\bf 25} (2005), 601--610.

\bibitem{X1}
{\sc X. Xu,} Stable-range approach to the equation of nonstationary
transoic gas flows. {\it Quart. Appl. Math.} {\bf LXV} (2007),
529--547.

\bibitem{X2}
{\sc X.Xu,} Flag partial differential equations and representations
of Lie algebras, {\it Acta Appl Math} {\bf 102}, 249--280.

\bibitem{X3}
{\sc X.Xu,} Stable-range approach to short wave and
Khokhlov-Zabolotskaya equations, {\it Acta Appl Math} DOI 10.1007/s
10440-008-9306-3.

\bibitem{X-L-Z}
{\sc T. Xia, Z. Lv and H. Zhang,} Symbolic computation and new
families of exact soliton-like solutions of Konopelchenko-Dubrovsky
equations, {\it Chaos, Solitons and Fractals} {\bf 20} (2004),
561--566.

\bibitem{Zha1}
{\sc S. Zhang,} The periodic wave solutions for the
(2+1)-dimensional Konopelchenko-Dubrovsky equations, {\it Chaos,
Solitons and Fractals} {\bf 30} (2006), 1213--1220.

\bibitem{Zha2}
{\sc S. Zhang} Symbolic computation and new families of exact
non-travelling wave solutions of (2+1)-dimensional
Konopelchenko-Dubrovsky equations, {\it Chaos, Solitons and
Fractals} {\bf 31} (2007), 951--959.

\bibitem{Zhi}
{\sc H. Zhi,} Lie point symmetry and some new soliton-like solutions
of the Konopelchenko-Dubrovsky equations, {\it Appl. Math. Comput.}
{\bf 203} (2008), 931--936.

\bibitem{Z-X}
{\sc S. Zhang and T. Xia,} A generalized F-expansion method and new
exact solutions of Konopelchenko-Dubrovsky equations, {\it Appl.
Math. Comput.} {\bf 183} (2006), 1190--1200.




\end{thebibliography}
\end{document}